Title

# From biogeochemical exchange to market exchange: the impacts of blue carbon on coastal wetland science

Author

Fernando Javier Ruiz-Iglesias

Institut de Ciència i Tecnologia Ambientals, Universitat Autònoma de Barcelona (ICTA-UAB)

FernandoJavier.Ruiz@uab.cat

Abstract

While there is widespread agreement that markets for ecosystem services (MES) have transformed conservation, it is less clear whether they have transformed the practice of environmental science to meet market needs for stable commodities. We examine this further through the case of blue carbon. Putting marine ecosystems on the MES agenda, blue carbon makes the case to incorporate coastal wetlands into carbon markets to finance their conservation and mitigate climate change. Using a mixed methods approach combining bibliometric and Natural Language Processing (NLP) analyses of peer-reviewed coastal wetland literature, a qualitative review of highly cited publications, and semi-structured interviews with blue carbon scientists, we argue that blue carbon has reshaped coastal wetland science: broadly toward strategic research on conservation and restoration to address global climate change, and specifically by reframing biogeochemistry to meet market demands for stable carbon commodities. Measured by number and share of papers, the blue carbon field is growing quickly and outpacing other research. Its papers are disproportionately represented among the most cited coastal wetland publications, suggesting a growth in both scientific production and authority. Our results show the emergence of blue carbon has redirected biogeochemical research on coastal wetlands from dynamic cycles and processes to stored and preserved carbon, aligning research with the stable carbon metrics and commodities needed for carbon accounting and markets. This strongly suggests that coastal wetland scientists, although skeptical of carbon offsets, are shifting their work to align with the needs of market and policy frameworks in well-intentioned efforts to promote coastal wetland conservation and restoration. If this continues, alternatives to market-based blue carbon policies are unlikely to emerge, making scientists' own doubts about their effectiveness a serious cause for concern.

## 1. Introduction

The ecosystem services framework was intended to convey the importance of the environment for the well-being of humans to policy makers and others who did not subscribe to environmentalist views. The central goal was not just to make nature *visible,* but to make nature *economically* visible in order to justify its conservation (Odum, 1969; Robertson, 2006). A germinal example of this is Costanza et al. (1997), which estimated the global monetary value of the world's ecosystem services and natural capital to be $US 33 trillion per year (Costanza et al., 1997). But while ecosystem services use monetary value as a lingua franca to convey the worth of the environment, its earlier proponents did not intend to create a market in which ecosystem services were bought and sold (Dempsey and Robertson, 2012; Gómez-Baggethun and Ruiz-Pérez, 2011). Even Costanza et al. argue that although global ecosystem services are worth $US 33 trillion per year, “many ecosystem services are literally irreplaceable” (Costanza et al., 1997). Despite these early intentions, monetary valuation—putting a price tag on nature—paved the way for markets in ecosystem services: even if irreplaceable, ecosystem services were made exchangeable (sold in markets) (Dempsey and Robertson, 2012; Gómez-Baggethun and Ruiz-Pérez, 2011; Lave and Doyle, 2021: 56). Markets for ecosystem services (MES) have become almost ubiquitous in contemporary conservation efforts.

There is widespread agreement that the ecosystem services framework has transformed conservation by shifting environmentalists’ focus away from confronting systemic drivers of environmental degradation to working within an explicitly capitalist framework (Robertson 2006, Dempsey 2016, Dempsey and Suarez 2016, Suarez, 2023, 2026). What is less clear is how ecosystem services have transformed the practice of environmental science (but see Lave, 2011, 2012; Randalls, 2011) to meet MES’s need for simple, stable and predictable proxies (Lave and Doyle, 2021; Robertson, 2006). In this paper, we use the case of blue carbon to examine this question in more depth through a bibliometric and Natural Language Processing (NLP) analysis of coastal wetland literature from 1970 to March 2026, a qualitative review of highly cited publications, and semi-structured interviews with scientists working on blue carbon. We argue that blue carbon has shifted the major focus of some coastal wetland research toward the needs of carbon accounting and markets for credible and robust commodities.

## 2. Blue carbon

Blue carbon refers to the carbon captured and stored in coastal wetlands—mangroves, seagrasses, and salt marshes. It is a proposed nature-based solution to climate change that posits conserving and restoring these ecosystems to (1) preserve large organic carbon deposits in their sediments (which, if degraded, can release additional $CO_2$) and (2) capture and store additional $CO_2$ from the atmosphere. Blue carbon builds from the idea that coastal wetland carbon fluxes are “amenable to management” (IPCC, 2019), making coastal wetland carbon a governable object through the co-production of science and policy (Thoni and Rummukainen, 2025). The report that introduced the term and the first peer-reviewed blue carbon paper illustrate this science-policy nexus, the ecosystem service rationale, and the field’s predominantly market-based policy framing.

The term blue carbon was coined in the report *Blue Carbon: The Role of Healthy Oceans in Binding Carbon* (Nellemann et al., 2009)*,* which was prepared by multiple United Nations Agencies as well as scientists from universities, the International Union for the Conservation of Nature (IUCN), and the Spanish National Research Council (CSIC). The Report was published in 2009 shortly before the United Nations Climate Change Conference in Copenhagen (COP-15) which, among other topics, had placed the Reducing Emissions from Deforestation and Forest Degradation (REDD) mechanism high on the negotiation agenda. The report aimed to inform this negotiation by laying out scientific case and policy proposals for a REDD-like mechanism for the oceans. To this end, it provided evidence supporting the role of marine ecosystems, especially coastal wetlands, in burying carbon, and proposed a list of policy recommendations to incorporate marine ecosystems into existing climate governance regimes, including carbon offsets. In other words, the goal of this original report on blue carbon was to put marine ecosystems on the ecosystem services agenda, making the case to conserve and restore coastal wetlands as a climate change mitigation strategy.

Blue carbon predominantly appeared in technical reports, policy briefs, and news articles until 2013 when peer-reviewed papers began to outpace other publications and communications (Quevedo et al., 2023). The first peer-reviewed article on blue carbon was published in 2011 (Mcleod et al., 2011). This article, aptly named *A blueprint for blue carbon*, explained the importance of protecting coastal wetlands (which they refer to as vegetated coastal ecosystems) because of the valuable ecosystem services they provide. In particular, coastal ecosystems' ability to sequester carbon more efficiently than terrestrial forests made a compelling case for their protection, and also for the further development of scientific knowledge about blue carbon:

> "Although their global area is one to two orders of magnitude smaller than that of terrestrial forests, the contribution of vegetated coastal [ecosystems] per unit area to long-term [carbon] sequestration is much greater […]. Despite [their] value in sequestering carbon, and the other goods and services they provide, these systems are being lost at critical rates and action is urgently needed to prevent further degradation and loss. Recognition of the carbon sequestration value of vegetated coastal ecosystems provides a strong argument for their protection and restoration; however, it is necessary to improve scientific understanding of the underlying mechanisms that control carbon sequestration in these ecosystems" (Mcleod et al., 2011).

The initial report (Nellemann et al., 2009) and peer-reviewed publication (Mcleod et al., 2011) outlined the task of the emerging blue carbon scientific field: to integrate coastal wetlands into the existing climate governance frameworks, including carbon offset markets. Many scientists have answered the call: blue carbon publications have grown from the initial report in 2009, and the initial peer-reviewed article in 2011, to 772 peer-reviewed articles as well as over 200 technical reports, book chapters, and dissertation theses by 2021 (Quevedo et al., 2023). There are at least two Blue Carbon Labs, specifically dedicated to blue carbon research – one in the University of Cadiz (Spain) and the other at Deakin University (Australia). In addition, international coalitions (e.g., the Blue Carbon Initiative) organize international blue carbon science and policy efforts.

That widespread institutional uptake has been accompanied by a market-based policy orientation (Thoni and Rummukainen, 2025) common in much of the ecosystem services paradigm (Suarez 2026). This is illustrated by publications that call for "capitalizing on the global financial interest of blue carbon" (Friess et al., 2022), or that aim to "operationalize marketable blue carbon" (Macreadie et al., 2022).

Operationalizing blue carbon or any other Markets for Ecosystems Services (MES) requires the production of a stable commodity; this comes at a trade-off as "ecosystems are characterized by a lack of the very same simplicity, stability, and predictability on which functional markets depend" (Lave and Doyle, 2021: 18). Therefore, scientists involved in operationalizing blue carbon are not only tasked with providing an understanding of coastal wetland carbon cycling, but with producing the stable commodities the market requires, a fact at least some blue carbon scientists are explicitly aware of:

> "Simplifying methodologies should help facilitate cost reductions and efficient project development […] and move the carbon market forward." (Kuwae et al., 2022)

Taken together, the understanding of coastal wetland carbon fluxes as manageable, the call for scientists to clarify the mechanisms of carbon sequestration, and the market-based policy orientation provide the context for understanding the changes in coastal wetland science discussed in the following sections.

## 3. Methods

To assess how and to what extent blue carbon has reshaped coastal wetland research, we used mixed methods: bibliometrics, Natural Language Processing (NLP), qualitative abstract reading, and semi-structured interviews. First, we produced an allotaxonograph (a map-like histogram of ranked pairs of words) to show which words distinguish the coastal wetlands literature before and after the introduction of blue carbon. Second, we used bibliometrics to quantify counts and shares of blue carbon papers over time. Third, we trained word embedding models to identify the words most similar to "carbon" before and after the introduction of blue carbon; these words were used to find and read representative abstracts (prioritized by citation count) to confirm usage and contextualize embedding results with concrete examples. Lastly, we complemented these findings with semi-structured interviews with seven of the forty most published blue carbon scientists, according to publication count in Scopus. Together, the methods present a layered argument: broad movements from the allotaxonograph, quantitative changes from publication counts, and qualitative shifts from embeddings, abstract reading, and interview data.

### 3.1 Selection of the Literature

We collected article and review titles, abstracts, journal names, and publication years from Scopus on 2 March 2026 using a broad search query to capture all literature related to coastal wetlands, irrespective of field or topic, from 1970 onward (see search query in supplementary material 1). Since the goal was to understand whether and how the emergence of the blue carbon scientific field reshaped coastal wetland research broadly, we refrained from using any specific key words such as “carbon cycling” or “ecology” that could narrow the selection and give a less comprehensive view.

We only considered publications written in English, which overlooked publications in other languages. Notably, publications in Chinese could be important because China has a comparable number of blue carbon publications to the top producing countries (the United States and Australia) (Zhong et al., 2023). Nevertheless, restricting the corpus to English ensured consistency in the text-analysis methods. Only articles and reviews were considered because abstracts were oftentimes unavailable for other kinds of publications. The search query resulted in 62,245 articles and reviews (hereafter publications or papers), from all sources, representative of Anglophone coastal wetland research broadly.

We classified the publications by their quartile according to the SJR-SCImago Journal and Country (SCImago, n.d.), and only considered the 31,362 publications printed in Quartile 1 (Q1) journals. This was done for two reasons. First, to ensure that the resulting publications in the database were of similar quality, or at least, held to the minimum quality control standards expected of Q1 journals. And second, to limit the size of the selected publications to lower the computational demands of the Natural Language Processing (NLP) analysis that used tools that could not handle the file size of the entire body of literature.

## 3.2 Bibliometrics

To quantify the extent to which blue carbon had emerged as a new area within coastal wetland research, we assigned each paper to one of three categories using simple keyword matching (grepl() in R). Papers included in the *blue carbon* category contained the phrase "blue carbon" in title, keywords or abstract; papers included in the *just carbon* category contained "carbon" but not "blue carbon" in title, keywords, or abstract; and the *other coastal wetland* category included all remaining papers. For each category we calculated annual counts and the share of total publications to track how both the number and proportion of papers changed over time.

## 3.3 Natural Language Processing (NLP)

To compare coastal wetland research before and after the emergence of blue carbon, we used two NLP approaches: an allotaxonograph and word embeddings. The data was split into two periods: 1970 to 2010 (before blue carbon) and 2011 to 2026 (after blue carbon), with 2011 chosen because it marks the first peer reviewed article using "blue carbon" in title, abstract, or keywords (Mcleod et al., 2011).

Our goal was to process the text in a way that minimized both alterations and sources of noise. For the allotaxonograph, titles and abstracts were combined for each paper and tokenized using the Natural Language Toolkit's (NLTK) word_tokenize. [1] Tokenization breaks text into individual word tokens for analysis. A small set of multi-word items that should be treated as single tokens were joined with underscores (e.g. "blue_carbon," "climate_change," "ecosystem_services," "salt_marsh," "tidal_marsh," and "machine_learning") so they would be counted as a single token. Standard English stopwords (e.g., "and", "the"), punctuation, and common publisher/license phrases found at the ends of abstracts were removed to reduce noise.

We used the py_allotax library to produce an allotaxonograph which compares the ranked word lists from the two periods and highlights which words most contribute to differences between them (Dodds et al., 2023; St-Onge et al., 2025).[2]. Put differently, the plot shows both how common a word is and how much it helps distinguish the "before" and "after" blue carbon literatures. The method ranks words by frequency in each period and computes a divergence contribution for each word; changes in rank are then scaled by a parameter (alpha) so that shifts in common words carry more weight than shifts in very rare words. For example, "mangrove" shifting from rank 2 to rank 1 is more informative than a one-rank change among a low-frequency word. We selected the alpha parameter by testing multiple values following the package guidance and chose alpha = 0.35 as producing meaningful outputs.[3] A summary of the scaled divergence rankings is found in a table adjacent to the figure.

For the word embeddings, the text was similarly tokenized with some exceptions. First, only abstracts were considered. Second, stop words were not removed but non-alphabetic tokens were removed, ensuring the word embedding model considered only text and produced interpretable results. And lastly, variants of specific keywords (e.g., “sequestering” and “sequestration”, “sulphur” and “sulfur”) were merged to a standard form.

---

[1] For more information, see documentation: https://www.nltk.org/api/nltk.tokenize.html
[2] See Dodds et al., 2023 for descriptions of the work leading up to the tools, and St-Onge et al., 2025 for a description of the tools themselves.
[3] For further information, see documentation: https://github.com/compstorylab/py-allotax

We trained two separate word2vec models for each time period using models.word2vec from the genism Python library.[4] Word2vec converts each word into a high-dimension vector based on its use in a text, where similar words are represented as nearby vectors. We then calculated the distances using the cosine similarity, the standard measure of distance in higher dimensions. To give a simple example, in the sentences “A cat has four legs. A dog has four legs,” the words “cat” and “dog” produce very nearby vectors because they appear in the same context. In this study, we identified the 15 words closest to “carbon” before (1970-2010) and after (2011-2026) blue carbon by ranking cosine similarity (0 -1), where higher values indicate greater closeness to “carbon.”

To ensure meaningful results, we kept only words that occurred at least 75 times because at this threshold the nearest neighbor lists were relatively stable while excluding most low frequency, less informative terms. The selected nearest neighbor words were then used to find and read representative abstracts (prioritized by citation count) to confirm usage and provide concrete examples of the shifts suggested by the embeddings. Concretely, we searched the data for each nearest neighbor term, sorted the matching publications by citation count, and then read the abstracts manually starting with the most cited.

### 3.4 Semi-structured interviews

[4] For further information, see documentation: https://radimrehurek.com/gensim/models/word2vec.html

To complement the textual analysis of the scientific literature on coastal wetlands, we conducted 7 semi-structured interviews with researchers working on blue carbon. The aim of the interviews was to contextualize any shifts in coastal wetlands science via the personal reflections and experiences of blue carbon scientists. Interviewees were selected from among the top 40 researchers based on the number of their publications included in Scopus that had the phrase "blue carbon" in the title, abstract, or keywords. Invitations were sent via email. Those who agreed to participate were provided with an information sheet and consent form outlining the nature and purpose of the study, as well as details about the voluntary nature of participation, confidentiality protocols, and data use.

Each interview lasted from 30 minutes to an hour and a half and was conducted online. A list of guiding questions was provided in advance. All interviews were recorded with the participant's consent. Transcripts were generated using Artificial Intelligence (AI) transcription tools. We then manually reviewed each transcript alongside the original video recording to ensure accuracy and correct any errors or ambiguities. We then analyzed the transcripts based on an initial set of codes that we expanded as we worked through the transcripts. Key codes included changes in research; personal career changes since blue carbon; market, policy & funding demands; personal motivations & opinions on carbon credits; institutional responses to climate change.

We obtained ethical approval from the Research Ethics Committee (CERec) of the Universitat Autònoma de Barcelona (reference code: UAB-CERec66), and written or verbal informed consent was obtained for anonymized interview quotes published in this article.

## 4. Shifts in Coastal Wetland Science

Our research revealed six major movements in coastal wetland science from before to after blue carbon: from basic to strategic research; from past to present and future framings; from marsh and seagrasses to mangroves; from plants to soils; and from dynamism to stability (Figure 1). Because after blue carbon there were almost three times more publications in our dataset (23,190) than before blue carbon (8,172), it is important to note that these shifts are relative. In other words, it is likely that what we are seeing is emerging research outpacing, not replacing, prior research.

In subsequent paragraphs, we provide a broad-brush analysis of the three of these shifts: from local to global, from basic to strategic research, and from past to present and future framings. A thorough analysis of these movements is beyond the scope of this paper; we therefore discuss these broad movements in relation to blue carbon, understanding that blue carbon is part of a larger shift in coastal wetland research. Two other movements (from plants to soils, and from dynamism to stability) will be analyzed in Section 6 in more depth alongside additional empirical evidence that focuses specifically on coastal wetland biogeochemistry. Plots showing publication count shifts from marsh (and to a lesser extent, seagrasses) to mangroves can be found in supplementary material 2.

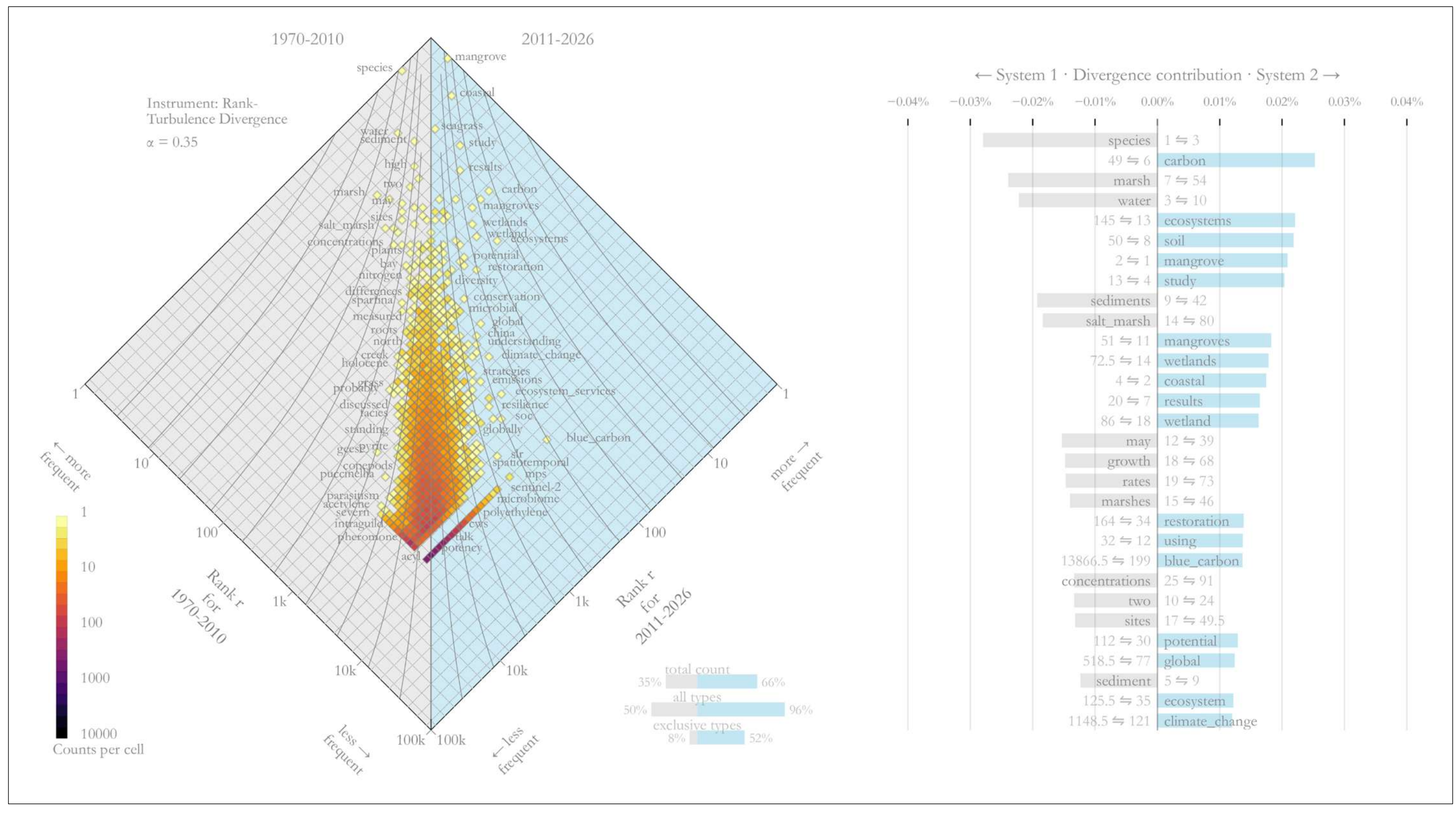

Figure 1**:** Allotaxonograph showing word frequency and which words contribute most to divergence between coastal wetlands literature before (1970-2010) and after (2011-2026) the first use of “blue carbon” in a peer-reviewed publication. Words near the top (e.g. mangrove, species, coastal) rank highly in both periods; words near the bottom (e.g. pheromone, intraguild) rank low. Words along the bottom right diagonal (e.g. potency, sentinel-2, and blue carbon) appear only after blue carbon. On the horizonal axis, distance from the center indicates relative change in rank: left = greater contribution to before blue carbon literature; right = greater contribution to after blue carbon. Following the same pattern, the table on the right lists the words contributing most to the divergence. Overall, words indicate a relative shift from basic to strategic research; from past to present and future framings; from marsh and seagrasses to mangroves; from plants to soil; and from dynamism to stability.

The words that suggest a shift from local to global in coastal wetland science include: "sites," "creek," and "bay" before blue carbon and "global," "globally," and "Sentinel-2" after blue carbon. The former set of words suggests research at specific localities, whereas the latter set indicates a global perspective and the use of larger-scale observations such as satellite imagery (Sentinel-2 is an Earth observation satellite). Although these shifts are occurring in coastal wetland research as a whole, they are likely driven in part by blue carbon. Early blue carbon papers emphasized the global significance of coastal wetland carbon stocks and estimated emissions from their degradation. These seminal studies include the estimate that mangrove deforestation generates emissions equivalent to 10% of global deforestation despite mangroves comprising only 0.7% of tropical forest area (Donato et al., 2011); a global estimate of seagrass carbon stocks (Fourqurean et al., 2012); and a global estimate of emissions from coastal wetland conversion and degradation (Pendleton et al., 2012). One interviewee was part of these early blue carbon efforts, and described the initial task of the emerging field to synthesize carbon data gathered for a variety of different reasons and localities in order to arrive at initial global estimates, "[early on] we said let's collate what we've got, and see how much carbon is stored" (interview 7). Additionally, these studies align with the increased prioritization of "climate change," whose governance is framed globally rather than locally (Miller, 2004; Turnhout et al., 2016).

The words that suggest a shift from basic to strategic research on coastal wetlands include "concentrations" and "species" before blue carbon and "conservation," "restoration," "ecosystem services," "nature-based," and "blue carbon" after blue carbon. The understanding deployed here between basic and strategic science is a simplifying heuristic, following previous science studies scholars that distinguish between autonomous and curiosity-driven basic science and reflexive and application-oriented strategic science (Gibbons et al., 1994; Irvine and Martin, 1984).[5] Specifically, strategic science is carried out with the expectation that it will form the basis of solutions to socially recognized problems (Rip, 2002). The words "conservation," "restoration," "ecosystem services," "nature-based," and "blue carbon" suggest strategic research to address socially recognized problems, including but not limited to climate change, sea-level rise and ecosystem loss. One interviewee reflected on this shift in relation to their personal career transition from plant physiology to blue carbon:

> "Years ago, most people focused on the seagrass' other ecological function – the nutrient uptake, and also the bacteria, but [now] the [focus is] how much [can] seagrass restoration enhance carbon sequestration." (interview 2)

These socially recognized problems appearing as distinguishing words in coastal wetland science after blue carbon also speak to a shift in research priority from past to present and future framed research. Before 2011, "facies" and "Holocene" appear as two of the distinguishing words. Facies are distinctive bodies of sediment that mark specific past and present environmental and geological conditions; the "Holocene" is the geological era of relative climactic stability of the last 11,700 years. A closer inspection of the most cited publications with these distinguishing words reveals that they are related to paleoecology, paleoenvironmental coastal geomorphology and reconstruction of past sea-level changes, ecosystem communities, and vegetation migrations (e.g. Chappell, 1983; Donnelly and Bertness, 2001; Gehrels, 1999; Scott and Medioli, 1978; Sloss et al., 2007). On the other hand, publications including "conservation" and "restoration" after 2011 stress the importance of coastal wetlands to address climate change and sea-level rise by reducing wave heights and preventing flooding and erosion (Narayan et al., 2016) and capturing and storing carbon to mitigate climate change (Alongi, 2012; Atwood et al., 2017; Macreadie et al., 2021; Murdiyarso et al., 2015).

[5] For secondary literature that explains the distinction in early science studies (Jasanoff, 2005; Pestre, 2003).

Analysis of the allotaxonograph thus shows a broad trend in coastal wetland research toward strategic science to address present and future global environmental change. The next sections will focus specifically on the growth of blue carbon within coastal wetland research and how it has affected coastal wetland biogeochemistry specifically.

## 5. The emergence of blue carbon

Scientific research on blue carbon has grown considerably, as demonstrated by the steadily increasing share of *blue carbon* publications in coastal wetland research from a single publication in 2011 (Mcleod et al., 2011) to 16% of yearly publications in 2026 (Figure 2). At the same time, the relative share of *other coastal wetland* publications has decreased from 84% of yearly publications in 2011 to 62% in 2026. The share of *just carbon* publications – those that contained "carbon" but not "blue carbon" in title keywords or abstract – has increased gradually from 16% of yearly publications in 2011 to 22% in 2026. Taken as a whole, the average yearly change in total publication share for *blue carbon*, *just carbon*, and *other coastal wetland* publications is +1.1%, +0.4%, and -1.5%, suggesting that the decreased share of *other coastal wetland* publications can be attributed to increased interest in carbon, particularly blue carbon.

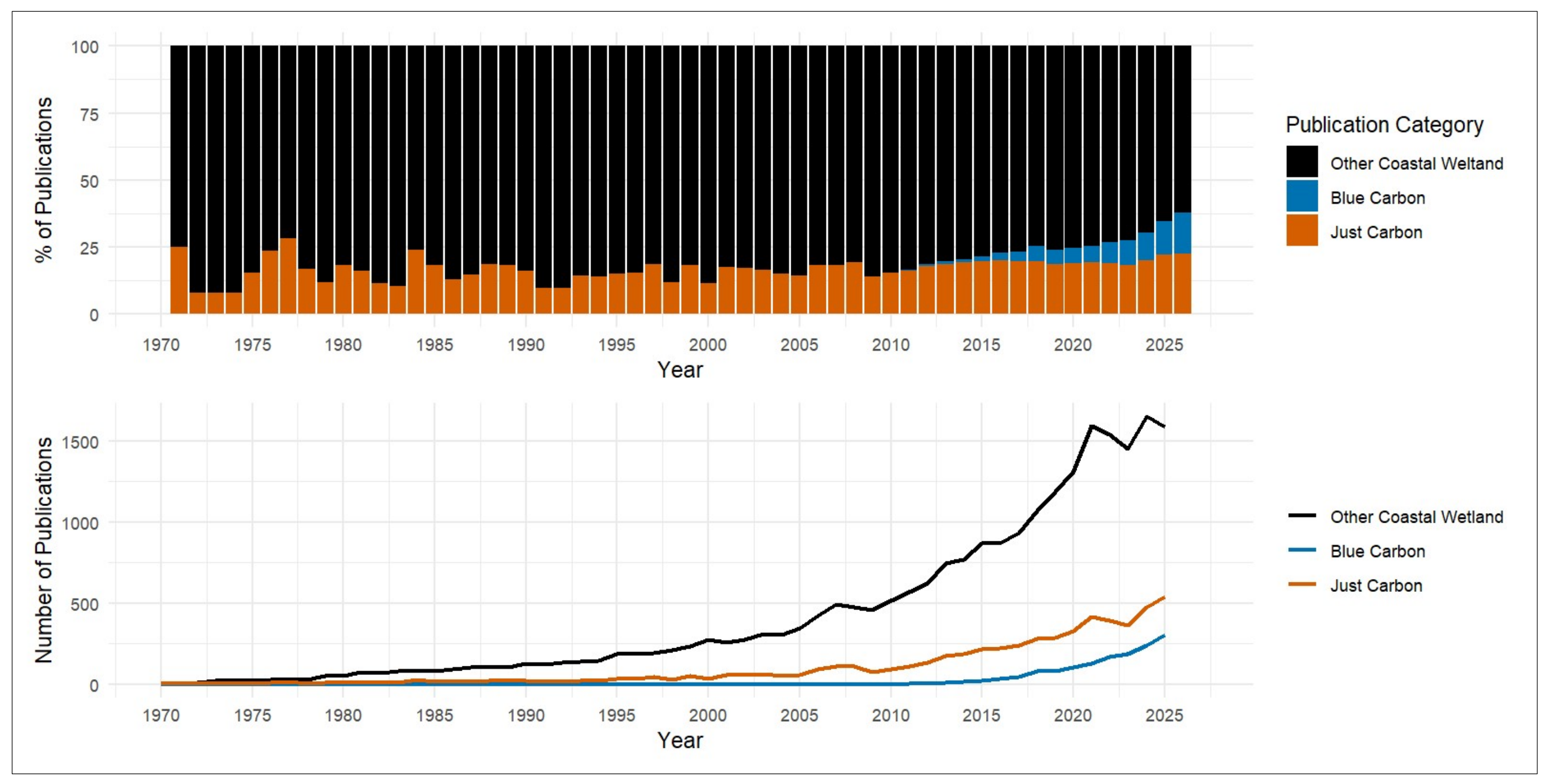

Figure 2: Yearly share (top) and yearly number (bottom) of publications for three categories: *blue carbon*, *just carbon*, and *other coastal wetland*. *Blue carbon* = papers with "blue carbon" in title, keywords or abstract; *just carbon* = papers with "carbon" but not "blue carbon" in those fields; *other coastal wetland* = all remaining papers. The bottom panel excludes 2026 because data were downloaded on 2 March 2026; the top panel includes yearly share for 2026 up to that date. *Blue carbon* share and number of publications increased most sharply; *just carbon* rose more gradually; *other coastal wetland* declined in share and appears to be plateauing in count.

In terms of absolute publication counts, publications in the *other coastal wetland* category have started to show signs of plateauing while *blue carbon* and *just carbon* publications increase. The yearly count of *blue carbon* publications has more than doubled and showed consistent growth over the last 5 years; *just carbon* publications oscillated but grew gradually; *other coastal wetland* publications have begun to show signs of stagnation (Figure 1). This shows a growing prioritization of coastal wetland carbon cycling over other topics, a finding that aligns with the reflections of a junior scientist whose masters and PhD coincided with the rapid growth of blue carbon:

> "Research was more […] into the ecology of seagrasses, mangroves and marshes, and less on the carbon cycling side. Now there's still the same or more research on ecology, but the carbon cycling has gained a lot of interest and traction." (interview 3)

The interest and traction described by the interviewee is hard to measure by looking only at publication counts as not each publication is equally cited. From 1970 until 2 March 2026 (the day the data was downloaded), there was a total of 1513; 5810; and 24,039; *blue carbon*, *just carbon*, and *other coastal wetland* publications that make up 5%, 18%, and 77% of the coastal wetland literature, respectively.

Although blue carbon publications only comprise 5% of the literature, it is disproportionately represented in the most cited publications. Among the 20 most cited coastal wetland publications, two are among the first blue carbon publications, that is, those that explicitly mention "blue carbon" in their titles, keywords, or abstracts (Fourqurean et al., 2012; Mcleod et al., 2011); three pre-date blue carbon but laid the foundational context for mangroves and salt marshes as important global carbon sinks (Chave et al., 2005; Chmura et al., 2003; Donato et al., 2011); and two discuss blue carbon as a climate change mitigation strategy in-text (Alongi, 2014; Duarte et al., 2013). Although these five latter publications do not contain "blue carbon" explicitly, they are described as foundational texts in bibliometric studies of blue carbon (e.g. Duarte De Paula Costa and Macreadie, 2022) In sum, although blue carbon represents just 5% of total coastal wetland publications, 35% of the 20 most cited publications could safely be considered part of the emerging blue carbon field.

Does this mean the dataset we assembled is missing substantial portions of the blue carbon literature? It might. The share and count of publications captured in the blue carbon category may be conservative because papers about blue carbon do not always contain that phrase in title, keywords, or abstracts. This is why previous bibliometric research typically included search queries beyond "blue carbon" and manually checked each publication for relevance (Jiang et al., 2022; Lai et al., 2022; Quevedo et al., 2023; Yin et al., 2023; Zhong et al., 2023).[6] However, the results could overestimate the impact of blue carbon on coastal wetland science if the phrase "blue carbon" is increasingly used for visibility even for research not directly related to coastal wetland carbon cycling. For example, the phrase "blue carbon ecosystems" is increasingly used as a collective label for mangroves, seagrasses and salt marshes, potentially replacing other collective terms such as "coastal wetlands" or "vegetated coastal ecosystems."

Despite these possible limitations, the disproportionate number of blue carbon articles among most cited publications in coastal wetland science suggests that it is growing not only in count and share, but also in scientific authority. Citation distributions are skewed following Zipf's law, meaning few top publications receive a disproportionate share of citations (Fedorowicz, 1982). Future work could undertake a more comprehensive study of citations and address the shortcomings of publication categorization here by using topic-based community detection (e.g. Latent Dirichlet Allocation) to classify publications based on topic instead of keywords.

Our data on publication counts and shares show that blue carbon has emerged as a relatively small but quickly growing share of coastal wetland research. Blue carbon publications make up a disproportionately large share of the 20 most cited publications in coastal wetland science. This raises the possibility of substantive shifts in a relatively short time. But is this actually the case? To answer that question, we turn to word embeddings and qualitative analysis of papers on coastal wetland biogeochemistry.

[6] That approach was not feasible for this article because of its expanded focus on blue carbon within the context of coastal wetlands research more broadly; this generated a much larger dataset that was not amenable to manual classification (which itself could have introduced interpretive bias).

## 6. Blue carbon and the reframing of biogeochemistry

Our analysis shows that biogeochemistry research has changed substantively with the emergence of blue carbon, as demonstrated by a word embedding analysis of the 15 nearest neighbors to “carbon” before and after blue carbon (Figure 3). As a reminder, word embeddings represent words quantitatively as vectors, allowing us to quantify the similarity of word meanings from context. The embedding results revealed three key changes pre- and post- blue carbon: the shift from exchange to storage, the decoupling of carbon and other elements, and the reframing of biogeochemical cycles around climate change. The following discussion demonstrates these patterns with concrete examples from the literature, selected by prioritizing publications that contain the embedding nearest neighbor terms.

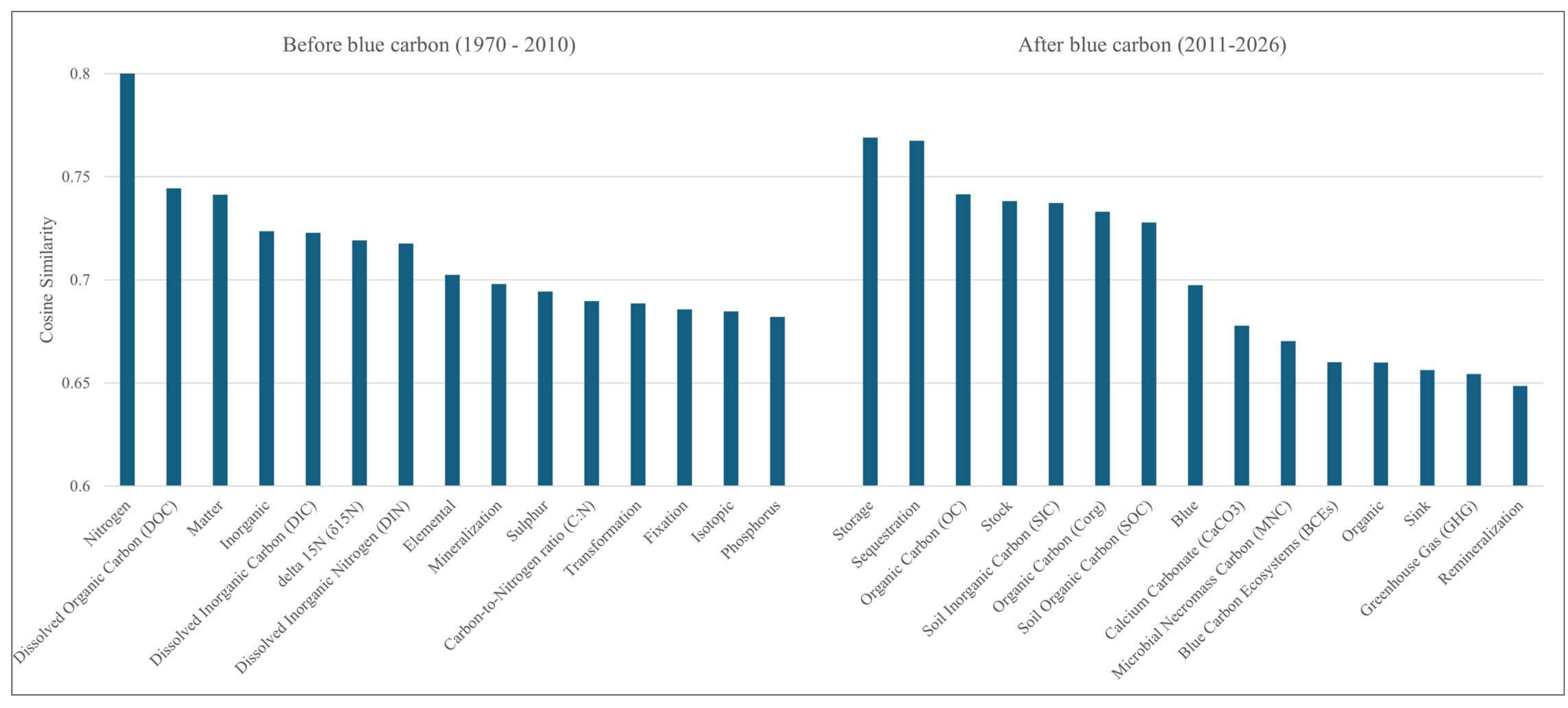

Figure 3: Bars show the cosine similarity values for the 15 words most similar to "carbon" before (1970-2010, left) and after (2011-2026, right) the first use of "blue carbon" in a peer-reviewed publication. Values range from 0 – 1; higher values indicate greater similarity to "carbon." Overall, the earlier period links "carbon" to other elements and dynamic biogeochemical processes, while the later period links it with stored organic carbon prone to greenhouse gas emissions.

Sitting at the boundary between land and oceans, coastal wetlands are in a constant state of flux and exchange. Vertically, matter moves from atmosphere, to plants, to soils and groundwater. Laterally, matter moves from land to coast and onwards to the ocean, through coastal wetlands. Dissolved Organic Carbon (DOC), Dissolved Inorganic Carbon (DIC), and Dissolved Inorganic Nitrogen (DIN) are transported by water both laterally and vertically, and appear among the nearest neighbors before blue carbon. Laterally, carbon is exported from mangroves to coasts, oceans, and estuaries both as dissolved inorganic carbon (Bouillon et al., 2008), and as dissolved organic carbon (Dittmar et al., 2006; Jaffé et al., 2004). Within coastal wetlands, dissolved inorganic matter can originate from a variety of sources, including from photochemical and biological processes that produce dissolved inorganic carbon from dissolved organic carbon (Miller and Moran, 1997; Miller and Zepp, 1995). Further, mangroves can also receive dissolved organic carbon from seagrass meadows (Bouillon et al., 2007), or from adjacent creeks during rising tides (Furukawa et al., 1997). Similarly, dissolved inorganic nitrogen is exchanged alongside organic matter in mangrove-estuary systems (Dittmar and Lara, 2001); from groundwater to salt marsh-estuary systems (Krest et al., 2000); or from land to coastal wetlands following heavy monsoon rains (Mukhopadhyay et al., 2006).

Importantly, isotopic carbon, nitrogen and phosphorus are measured to determine the origin and sources of matter in these exchanges, not carbon alone. For example, between connected river-coastal-oceanic systems (e.g. Cloern et al., 2002; Peterson et al., 1986; Zhang et al., 2007). In addition, carbon, nitrogen and phosphorus are measured to characterize the exchanges of organic matter between trophic levels in food webs (e.g. Bouillon et al., 2002; Davenport and Bax, 2002; Kwak and Zedler, 1997; Newell et al., 2012; Peterson et al., 1985; Peterson and Howarth, 1987; Wada et al., 1991). In their elemental form, the balance of these elements is key to understanding the mutually dependent relationships between plants and microbes. When vegetation dies, microbial communities transform nutrients from dead vegetation into sources of nitrogen, phosphorus, and sulphur that can be used by plants (Holguin et al., 2001). For this reason, there has been a substantial body of scientific research analyzing how plants respond to nitrogen and phosphorus inputs either experimentally (Feller, 1995; Feller et al., 2003; McKee et al., 2007) or as a result of pollution (Slomp and Van Cappellen, 2004).

The process by which microbial nutrients transform nutrients from dead vegetation into forms that can be used by plants is called mineralization. Mineralization depends, among other factors, on the balance between carbon and nitrogen. Hence, both "C:N" the abbreviation of carbon-to-nitrogen ratios and "mineralization" are among the nearest neighbors before blue carbon in Figure 3. But after blue carbon, there seems to be a decoupling of carbon and nitrogen (and possibly other nutrients). The trends in publications containing "carbon" and "nitrogen" in the title, keywords, or abstract indicate that, after the emergence of blue carbon, carbon and nitrogen are less frequently measured together (Figure 4). Before blue carbon, publication counts of carbon and nitrogen oscillate in tandem, suggesting the two are measured together. After blue carbon, carbon publications increase substantially whereas nitrogen publications increase only modestly. This indicates a decoupling of carbon and nitrogen in the literature with the emergence of blue carbon – a change noted by a senior scientist who experienced the shift as they moved into blue carbon research later in their career:

> "People don't look at the nitrogen, even if it's measured, and plays a major role in how the system works [...] a lot of the work is [...] just [carbon] stock and accumulation rates, not looking [at] how the system works.*"* (interview 1)

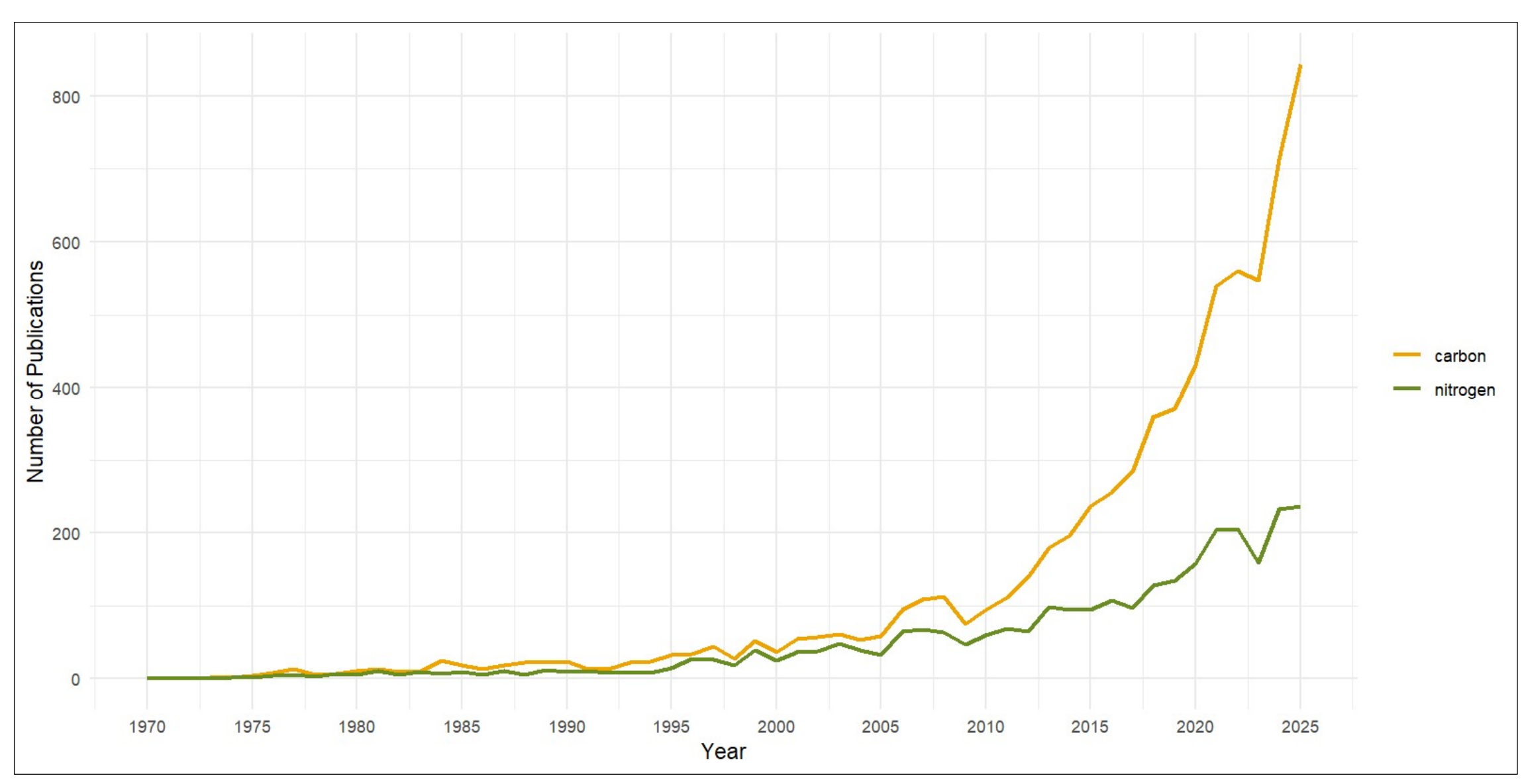

Figure 4: Yearly number of carbon and nitrogen publication. Carbon = papers with “carbon” in title, keywords or abstract; nitrogen = papers with “nitrogen” in those fields. Categories are not exclusive (papers can mention both). Before 2010, the two categories oscillate together; from 2010 onward, carbon publications rise sharply while nitrogen publications increase only modestly, indicating a decoupling of carbon and nitrogen in the literature with the emergence of blue carbon.

This is not to say that mineralization is no longer considered; instead, it has been reframed to fit the blue carbon framework via a stronger focus on remineralization, which is used more often in blue carbon publications as it aligns with blue carbon’s conservation framing. Studies using “remineralization” frame the process as a *loss* of stored carbon (Atwood et al., 2017; Kelleway et al., 2020; Trevathan-Tackett et al., 2017a, 2017b); they stress the susceptibility of carbon stocks to remineralization and the importance of conserving them. As one article states:

> “Widespread deforestation threatens the preservation of important [mangrove soil] carbon stock[s]. It is therefore imperative that global patterns in mangrove soil carbon stocks and their susceptibility to remineralization are understood” (Atwood et al., 2017).

Most of the organic carbon stored in coastal wetlands is found in their soils (Alongi, 2012; Fourqurean et al., 2012). This soil organic carbon is relevant to climate change because coastal wetland soils can preserve organic carbon for millennia due to anoxic, or oxygen-poor, conditions and other factors that inhibit decomposition and effectively

“lock up” carbon (Duarte et al., 2013; Mcleod et al., 2011; Pendleton et al., 2012).[7] The stored >30,000 Tg of carbon in coastal wetlands is primarily how the emerging blue carbon field urges coastal wetland conservation, because if released, this carbon could amount to substantial GHG emissions (Macreadie et al., 2021). For this reason, the nearest neighbors with the highest similarity scores after blue carbon include “sequestration,” “storage,” and “stock.” Stored organic carbon and its preservation is increasingly becoming the primary focus in the literature – a notable difference from the prior nearest neighbors of carbon which denote the dynamism and connectivity of coastal systems.

Importantly, our interview data confirms that these shifts in coastal wetland research stem from its alignment with carbon accounting and market-based policies that prefer stable quantities over dynamic flows. One interviewee summarized the shifts in research foci from dynamism to stability, and how coastal wetland science has changed from describing flows to quantifying soil carbon stocks:

> "In the past [… coastal wetland research] was more detailed [in] trying to explain how carbon flows through the different compartments, not only the soil, but from the biomass to the soil to the water column. There was more nuance. I think now there's more and more data on just one of the compartments like soil carbon and then there's only papers about soil carbon” (interview 3).

This raises the question of whether blue carbon science reflects how coastal systems work, or whether they reflect what market-based policy frameworks demand. One interviewee made clear that most of blue carbon science was collecting data on stocks and accumulation, deviating from descriptions of the carbon cycle,

> "A lot of the work that has been done in blue carbon... has been producing data of how much carbon and at which rate it is accumulated in this setting, not necessarily related to the actual carbon cycle." (interview 1)

[7] Other conditions include but are not limited to low plant tissue nitrogen and phosphorus concentrations, the allocation of a large fraction of biomass production to below-ground roots and rhizomes, and dense canopies of vegetation as well as root and rhizome networks that protect stored soil carbon from erosion (Duarte et al., 2013).

We are not claiming that coastal wetland scientists have forgotten how the carbon cycle works. Instead, what seems more likely is that they are responding to changing demands from policy and funding sources. As one interviewee explained it:

> "We [scientists] are talking about the details and the nuances, but then the funding goals and policy are asking how we can simplify accounting. So I don't conceptualize carbon cycling differently, but the deliverables that the funding [require] are changing." (interview 3).

Strikingly, the scientists we interviewed are responding to these demands in well-intentioned efforts to conserve and restore coastal ecosystems rather than out of hope that blue carbon projects might meaningfully address climate change. They did not believe that conserving and restoring coastal wetlands would solve climate change and understood that the institutional support for carbon crediting enables postponement of the "more thorny and complex issue of reduction in actual emissions from burning fossil fuels" (interview 7). But despite their skepticism about the efficacy of carbon crediting and their acknowledgement that other socioecological services provided by coastal wetlands – such as coastal protection and resource and food provision – were more important, nearly all interviewees supported blue carbon as a crucial opportunity to preserve coastal ecosystems. As one interviewee noted,

> "[Conservation/restoration of coastal wetlands for climate change mitigation] is not the best way. The only reason I'm really interested in it is that I think it does motivate conservation and restoration of coastal wetlands, but it is not a solution to global climate change" (interview 4).

Thus, the emerging field of blue carbon condenses the complexities of cycles into stable measurements needed to integrate coastal wetlands into carbon accounting and markets despite scientists' concerns about the biogeochemical credibility of that task. Because as it stands, at least from one interviewee's perspective, "if there's no carbon sequestration, there's no dollars" (interview 1).

## 7. Conclusions

Although just 5% of the coastal wetland literature, blue carbon punches well above its weight. Measured by number of papers and as a share of overall literature, the field is quickly growing, and its papers are disproportionately represented in the most cited publications. Our results suggest that the rising influence of blue carbon has led to substantive shifts in coastal wetland research. Our allotaxonograph analysis showed three key movements since 2011 – from local to global, from basic to strategic science, and from past to present and future framing – that appear to be driven in part by blue carbon's focus on conserving and restoring coastal wetlands to mitigate the present and future effects of global climate change.

To more carefully examine whether blue carbon is in fact creating significant changes in coastal wetland research, we conducted word embedding analysis and interviews with blue carbon scientists. This work showed that the emergence of blue carbon has redirected a substantial portion of biogeochemical research on coastal wetlands from dynamic cycles and processes to stable and preserved carbon, aligning research with the stable carbon metrics needed for carbon accounting and markets.

Taken together, our results show that scientists have increasingly shifted their analysis to align with the needs of market and policy frameworks in well-intentioned efforts to promote coastal wetland conservation and restoration. To our knowledge, we are the first to document this specific shift in peer reviewed literature in relation to ecosystem services, but the history of science is rife with examples of science shifting substantively in relation to political economic forces. Dominique Pestre refers to this as science regimes: Galileo, for example, produced very different science depending on his patron (Pestre, 2003). Mirowski has described this in relation to contemporary science, and how the privatization of universities and intellectual property has shaped research agendas (Mirowski, 2011).

Is this good or bad? Well, that depends. On the one hand, the end goal of blue carbon – the conservation and restoration of coastal wetlands for their climactic and other benefits – is hard to argue with. On the other hand, any benefits are contingent on the effectiveness of subsequent policies. Meanwhile, we show that such policies, effective or otherwise, have already changed scientific production. As research continues to be produced within the blue carbon framework, the likelihood of a genuinely different paradigm that moves beyond ecosystem services and markets emerging diminishes. Therefore, future work should evaluate the effectiveness of blue carbon policies and carbon offsets, since continued production of science under the current framework will likely only strengthen it and dimmish the possibility of any alternative.